\pdfoutput=1
\documentclass[
a4paper, %A4 paper, for convenience in Australia.
pdftex, %LaTeX output profile (e.g. pdftex/dvips/dvipdfm)
aps,
pre,
showpacs,
showkeys,
reprint,
groupedaddress,
amsmath,
amssymb,
floats
]{revtex4-1}

\usepackage{graphicx} % For including the figures.
\usepackage{enumerate} % More powerful enumeration, used for Appendix A
\usepackage{bm} % Easy bold math.

\newcommand{\texorpdfstring}[2]{#1}
\usepackage{ifoption} % For detecting output mode.
\IfClassOption{revtex4-1}{xetex}{
	\usepackage[bookmarks,bookmarksopen,bookmarksnumbered,bookmarksopenlevel=1]{hyperref}
}

\IfClassOption{revtex4-1}{pdftex}{
	\usepackage[bookmarks,bookmarksopen,bookmarksnumbered,bookmarksopenlevel=1]{hyperref}
}

\IfClassOption{revtex4-1}{dvips}{
	\usepackage[breaklinks]{hyperref}
}

\IfClassOption{revtex4-1}{dvipdfm}{
	\usepackage{hyperref}
	\DeclareGraphicsExtensions{.eps}
}

\newcommand{\nomenclature}[3][]{}

\newcommand{\norm}[1]{\left| \left| {#1} \right| \right|}
\newcommand{\N}{\mathbb{N}}

\begin{document}
\title{Evaluating transport in irregular pore networks}
\thanks{Published in Phys. Rev. E 86, 011112 (2012)}
\author{Dimitri A. Klimenko}
\email{dimitri.klimenko@uqconnect.edu.au}
\author{Kamel Hooman}
\author{Alexander Y. Klimenko}
\affiliation{School of Mechanical and Mining Engineering,
The University of Queensland, Qld 4072, Australia}
\date{\today}

\begin{abstract}
A general approach for investigating transport phenomena in
porous media is presented.
This approach has the capacity to represent various kinds of irregularity
in porous media without the need for excessive detail or computational effort.
The overall method combines a generalized Effective Medium Approximation (EMA)
with a macroscopic continuum model in order to derive a transport equation
with explicit analytical expressions for the transport coefficients.
The proposed form of the EMA is an anisotropic and heterogeneous extension
of Kirkpatrick's EMA [Rev. Mod. Phys. 45, 574 (1973)]
which allows the overall model to account for microscopic
alterations in connectivity
(with the locations of the pores and the orientation and length of the throat)
as well as macroscopic variations in transport properties.
A comparison to numerical results for randomly generated networks with
different properties is given, indicating the potential for this methodology
to handle cases that would pose significant difficulties to many other
analytical models.
\end{abstract}

\pacs{05.60.Cd, 81.05.Rm}
\keywords{porous media, transport,
pore network modeling, effective medium,
irregular pore network, continuum approximation, diffusion}

\maketitle

\section{Introduction}
There are a large number of applications for both natural and designed
porous materials including those in fuel cell technology \cite{tamayol11},
physiological transport phenomena \cite{truskey09},
and heat exchangers \cite{odabaee12, jaeger10}.
By natural porous media, we mean structures with random features;
like irregular pores in view of their sizes and shapes as well as
their interconnections.
On the other extreme, the term designed porous media \cite{bejan04}
refers to purposefully constructed porous structures where pores are designed
to perform functions under certain constraints;
with heat exchanger tube bundles in cross flow as a
very good example \cite{bejan93,hooman10}.
While porosity can be very high for designed porous materials,
natural porous materials typically have much lower
values for porosity \cite{bejan04}.

Although it is typically the macroscopic transport properties of
either a designed or a natural porous medium in which we are interested, 
these properties are irrevocably tied to the structure of the medium on a
microscopic scale.
However, due to the complexity of the pore-scale geometry,
full analytical results at a microscopic scale are nearly impossible,
while numerical simulation tends to be highly computationally intensive
and require excessive detail.
The approaches of volume averaging \cite{whitaker99} and conditional averaging
\cite{klimenko12} provide simplified analytical equations, but these equations
remain difficult to associate with the microstructure of the porous medium,
especially its connectivity.

A convenient approach that allows at least some of these microscopic properties
to be accounted for is pore network modeling, which represents the pore space
as a large number of pores connected together by narrow throats with simplified
geometry for both.
With the use of numerical simulations on networks reconstructed from
real-world media, realistic predictive estimates of transport
coefficients have been made \cite{kharusi07,tamayol11-pre}.
However, such reconstruction processes are relatively difficult, and
network simulations can be quite computationally intensive for very large
numbers of pores. Analytical models with suitable averaging remain highly
desirable for the purposes of efficiency and robustness.

Unfortunately, there are relatively few analytical approaches to modeling of
porous media that take the network microstructure into account.
One such approach is the continuous-time random walk (CTRW) method
\cite{sahimi12,fedotov07,berkowitz06}; however, apart from some
recent efforts \cite{sahimi12}, the current CTRW formulation usually
requires its key parameter to be fitted to experimental data.
Another group of approaches is the effective medium
approximations (EMAs), which can be applied to stochastic networks,
replacing them with deterministic ones on the condition that the average
of the local fluctuations must be zero.
Typically, an effective medium approximation is applied to a
regular network structure \cite{kirkpatrick73},
and then transport coefficients are calculated
by assuming the existence of a smooth macroscopic field,
and calculating local fluxes and hence transport coefficients based on this
Smooth Field Assumption (SFA) \cite{burganos87,jackson77}.
In general, effective medium approximations are reasonably accurate
as long as the medium is not near its percolation limit \cite{koplik81}.

The limitations of standard EMAs are demonstrated by 
network structures reconstructed from three-dimensional (3D)
imaging of samples of real-world porous media \cite{kharusi07}.
It is clear that regular lattices are not representative of the complex,
irregular nature of porous media, 
and although they can often be tuned in order to match empirical data,
the parameters in this tuning are typically just fudge factors.
In real-world media, the pores and throats vary at macroscopic scales,
the throat properties are often anisotropic, and the coordination number
exhibits significant variance.
Some extensions of the effective medium theory have been proposed that
deal with one or two of these issues
(anisotropic EMA \cite{bernasconi74,toledo92,sahimi00},
effective throat area \cite{vrettos89},
variable coordination number \cite{sahimi97}),
but an effective medium theory which deals with truly irregular networks
of the kind we expect from natural porous media seems to be absent from
the literature.

In this paper, we propose a generalized single-bond EMA which does not
require a recurring network topology,
and can have both anisotropy and large-scale variation.
This is combined with an explicit derivation of the transport equations
for the effective medium network,
in the form of a diffusion-like partial differential equation.

Section~\ref{sec:model} describes a general pore-scale model of transport
within porous media, in which a porous medium is represented as an irregular
pore-throat network which does not conform to any lattice pattern.
Pore-scale transport is then modeled using general transport equations
based on conservation principles.

Section~\ref{sec:ema} presents a generalized EMA, which is used to account for
microscopic variations due to poor connectivity within the medium.
The result of this approximation is a discrete pore network, but one
with deterministic throat conductances.

Section~\ref{sec:continuum} describes the continuum approximation used in 
order to derive an analytical expression for the transport coefficients in the
medium. 

Section~\ref{sec:results} shows a comparison of the analytical
method of Sections \ref{sec:ema} and \ref{sec:continuum} 
against numerical simulations for randomly generated pore networks.

\section{Pore-scale model} \label{sec:model}
In order to achieve a tractable representation of a porous medium at the
microscopic level, we begin our analysis with a stochastic pore-throat network model,
representing the medium as a network of pores connected together by
narrow throats.
Although the network model we consider is nonetheless a simplification of the
structure of real porous media, we avoid many shortcomings of network models by
considering a general class of stochastic networks.
Using a stochastic model allows for realistic modeling of porous media by only
using the important data about the statistical properties of a medium at a
microscopic level, such as types and distributions of connections and pores.
The stochastic properties then serve to represent uncertainty about the exact
microstructure while taking such information into account.
Moreover, the types of networks considered are capable of representing porous
media with various kinds of irregularity, including macroscopic heterogeneity,
significant variation in co-ordination numbers, and anisotropy.

Based on a simplification of real media, we consider the pore throats to have
negligible storage capacity (typically volume), so that accumulation of the
transported quantity occurs within the pores.
However, the throats are the primary consideration in determining the 
throughput within the medium, because overall transport
is limited by the constricted nature of the throats.
In particular, if there were no throats at all, the medium would consist
only of isolated pores, and no transport would take place.
As such, transport is primarily modeled within the throats, and at the
pore/throat interfaces.

We represent the medium by a network of $N$
\nomenclature[AN]{$N$}{Number of pores in the network}
pores, which
we enumerate with integers $1, \ldots, N$, where $N$ must be large in order
for a distinction to emerge between
macroscopic and microscopic properties of the medium.
A single pore within this network, labeled by its number $i$,
\nomenclature[ii]{$i, j$}{Pore number}
is then described by its location in $n$-dimensional space,
\nomenclature[bn]{$n$}{Number of spatial dimensions}
$\bm{x}_i = (x_i^1, \ldots, x_i^n)$,
\nomenclature[bx]{$\bm{x}$}{Pore position in space}
\nomenclature[p1]{$1, 2$}{Vector components}
which we take to be the position of its center,
and a vector of additional properties $\bm{\xi_i}$
\nomenclature[gxi]{$\bm{\xi}$}{Vector of additional properties}
(e.g. pore volume).
For our purposes, we combine the physical coordinates and the additional
properties into the extended vector $\bm{y}_i = (\bm{x}_i; \bm{\xi}_i)$
\nomenclature[by]{$\bm{y},~\bm{Y}$}{Extended position/property vector}
taking values in some vector space $S$,
\nomenclature[AS]{$S$}{Extended space (physical coordinates and properties)}
which we will call ``extended space''.

In order to describe the statistical uncertainty that is inherent in modeling
the complex and disordered structure of a porous medium,
the extended coordinate vectors are considered to be random variables
$\bm{Y}_1, \bm{Y}_2, \ldots, \bm{Y}_N$, corresponding to the positions
and properties of pores sampled from the medium without replacement.
As such, any individual $\bm{Y}_i$ has an identical marginal PDF 
(Probability Density Function), $f_{\bm{Y}} (\bm{y})$.
\nomenclature[bf]{$f$}{Probability density function}
For notational convenience, whenever the random variables considered are
evident from the arguments, the subscripts will be dropped - e.g. $f(\bm{y})$ 
for the single-pore distribution.

Throughout this network, there are throats connecting pairs of pores - these
are thin links through which transport between pores can occur. 
In general, like the pores, these throats could have various properties
associated with them.
However, for the purposes of the pore-throat network model, these properties
are combined into a single value which we call the conductance of the throat,
by analogy to electric circuits.
For two pores $i$ and $j$ the conductance of a throat connecting them
is the value $g_{ij}$,
\nomenclature[bg]{$g,~G$}{Throat conductance}
which describes the ease of transport through the
throat - a value of $0$ would mean there is no direct link,
while a value of $\infty$ would effectively combine the two pores into a
single pore.
Moreover, we do not consider the conductance of an individual throat to
depend upon the direction of transport, i.e. $g_{ij} = g_{ji}$.

As with the pores, we consider this conductance to be a stochastic variable,
$G_{ij}$,
since individual throats in the network could have a variety of values
for conductance.
It is reasonable to assume that this conductance depends primarily on
the properties of the throat itself, and on the pore-throat interfaces
on either side.
Of course, we can expect correlation between the throat and pore
properties - for example, larger pores would generally have larger
throats.
With this in mind, we express the distribution for conductances
for any given throat by
\begin{equation}
	f_{G_{ij}}(g) = f(g \mid \bm{y}_i, \bm{y}_j) ,
\end{equation}
which accounts for the effect of pore properties and locations on the
throat conductance,
as well as for independent randomness within the throats themselves.
It is important to note that throat length is also accounted for in this
representation, since the physical coordinates of the pores are included in
$\bm{y}_i$ and $\bm{y}_j$.

In a realistic porous medium, it is clear that the vast majority of
pairs of pores do not have any throat connecting them at all, especially
when the pores are spatially distant.
Considering this, we let $\Psi(\bm{y}_i, \bm{y}_j)$ be the probability
that a throat between pores with extended coordinates $\bm{y}_i, \bm{y}_j$
is present,
\nomenclature[gPsi]{$\Psi$}{Variable probability of throat presence}
and so the conductance distribution for the throat can be written as
\begin{align}
	f(g \mid \bm{y}_i, \bm{y}_j) = 
		\Psi(\bm{y}_i, \bm{y}_j) 
			f^*(g \mid \bm{y}_i, \bm{y}_j) \notag \\
		+	[1-\Psi(\bm{y}_i, \bm{y}_j)] \delta (g) ,
\label{eq:dist_g}
\end{align}
where $f^*$ is the distribution over nonzero values of conductance (i.e.
conditional upon $G > 0$),
\nomenclature[pstar]{$*$}{Conditional on the presence of a throat, i.e. $G > 0$}
and $\delta$ is the Dirac delta function.
\nomenclature[gdelta]{$\delta$}{Dirac delta function}

This stochastic network model is a very general type of random graph,
with the addition of random edge weights - the conductance values
representing the transport properties of the throats.
Of the many classes of random graph studied in the literature,
the closest well-known one is the 
Random Geometric Graph (RGG) \cite{dall02} due to the spatial
nature of the node parameters and localized connectivity.
Another approach is fitness-based connectivity \cite{caldarelli04},
but this concept is targeted towards the generation of scale-free
networks and hence is not used here.

Having defined the throat conductance, we can derive a general transport
equation valid for a variety of different transport phenomena,
including diffusion, conduction heat transfer, flow of electrical current,
transport across biological membranes, and linear cases of fluid flow
(e.g. incompressible Newtonian Stokes flow).
We consider inter-pore transport to be driven by a ``potential difference''
between two pores, while the transport phenomenon represents the tendency
for these potentials to equilibrate.

In this paper, we will use the terminology of molecular diffusion,
and so we identify the potential of a pore as the
concentration of the substance within the pore, $c_i = m_i / V_i$,
\nomenclature[bc]{$c$}{Amount of substance concentration}
where $m_i$ is the amount of the substance in the pore,
\nomenclature[bm]{$m$}{Amount of substance}
and $V_i$ is the volume of the pore.
\nomenclature[AV]{$V$}{Pore volume}
There is a net flow of particles from pores with higher
concentrations to those with lower concentrations,
resulting in a flow rate through any given throat
(from pore $i$ to pore $j$) defined by
\nomenclature[bq]{$q$}{Amount of substance flow rate}
\begin{equation}
	q_{ij} = -q_{ji} = g_{ij} (c_i - c_j) .
\end{equation}
With the use of a conservation law within the $i$-th pore, we can
finally derive the \textit{master equation},
\begin{equation}
	\frac{d m_i}{dt} = \sum_{j=1}^N q_{ji} = \sum_{j=1}^N g_{ij} (c_j - c_i) ,
	\label{eq:master}
\end{equation}
which is a system of $N$ differential equations that describes transport
within the medium at the pore scale.
As noted previously, this equation is applicable to a wide range of transport
processes;
expressions for throat conductance $g$ in terms of 
the geometric properties of the throat and
the properties of the materials or substances involved
are given in Appendix~\ref{app:trans} for several of these transport
phenomena.

\section{Nonuniform effective medium approximation} \label{sec:ema}
If the porous medium is very well-connected, global transport can
be directly approximated using the continuum approximation described 
in Section~\ref{sec:continuum}
using the mean or expected value for the conductance of each throat,
\nomenclature[zb]{Overbar (e.g. $\bar{g}$)}{Mean value}
\begin{equation}
	\bar{g} (\bm{y}_i, \bm{y}_j ) =
		\int_0^\infty g f(g \mid \bm{y}_i, \bm{y}_j) 
		\, dg .
\end{equation}
However, in general, the media we consider do not meet this requirement.
In order to resolve this, we apply a nonuniform effective medium approximation
that transforms networks with intermittent throats or significant
variations in conductance into well-connected ones.

The importance of connectedness within a pore network is highlighted
by many results in the area of percolation theory \cite{bedrikovetsky93}.
In particular, a very general result for such systems is that
in the limit $N \to \infty$, there is a critical level of connectedness
for the system which is called the ``percolation threshold'' - 
below this level, the overall network collapses into small, disjoint
clusters, and hence there is no transport at the macroscopic level
\cite{hunt09}.

In general, it is extremely difficult to derive the percolation threshold,
or transport behavior in the vicinity of the threshold, for all but the
simplest of network structures.
This is especially pertinent to the proposed network model, since it has
an irregular structure and correlations between the pores and throats.
However, a simple analysis of the proposed model can identify some
important structural properties.
Using $\Psi$, the probability that a throat is present,
we can derive the expected coordination number of a pore
with extended coordinates $\bm{y}$ as \cite{caldarelli04}
\nomenclature[bz]{$z$}{Expected coordination number / degree}
\nomenclature[pc]{$\circ$}{Dummy variable of integration}
\begin{equation}
\label{eq:z}
	z(\bm{y}) = (N - 1) \int_S f(\bm{y}^{\circ})
		\Psi(\bm{y}, \bm{y}^{\circ})
		d \bm{y}^{\circ}
\end{equation}

The value of $z$ is crucial to overall transport in the medium - if this
value is high, the network is well-connected and using the mean conductance
for every bond should prove to be a reasonable approximation.
Unlike with $N$, we cannot simply assume that $z$ is large -
pore networks extracted from 3D images of various porous media
\cite{dong09,kharusi07,bakke03,sheppard2005,raoush05,jiang2007}
have usually had average coordination numbers between 2 and 7.

To illustrate the effects of low values of $z$, consider the simple
case of a regular lattice pattern,
in which every throat independently has probability $p$ of being present 
($g = g_0$) and probability $1-p$ of being absent ($g=0$).
\nomenclature[bp]{$p$}{Constant probability of throat presence}
Using a single-bond EMA, Kirkpatrick \cite{kirkpatrick73}
has shown that the global transport properties of such a network are roughly
the same as those of an identical lattice pattern in which every throat
is present, with conductance
\footnote{Note that in our terminology $z$ is the mean coordination number of
the stochastic network, while Kirkpatrick's $z$ is equivalent to our
$\tilde{z}$.
In this simple example $\tilde{z} = z/p$.}
\nomenclature[zt]{Tilde (e.g. $\tilde{g}$)}{Value in the effective medium}
\begin{equation} \label{eq:g_simp}
	\tilde{g} = g_0 p \frac{z - 2}{z - 2p} .
\end{equation}
As $p \to 1$ or $z \to \infty$, Eq.~(\ref{eq:g_simp}) approaches the mean
value approximation, $\bar{g} = g_0 p$.
However, in cases where $p$ and $z$ are small, it is clear that the mean
value approximation is wrong by a significant margin, and hence a better
approximation is required.

Unfortunately, in general, single-bond EMA cannot account for behavior near
the percolation threshold of the system, because of a failure to account
for the effects of correlations between the throats in the system
\cite{koplik81} - in the terminology of percolation theory, the correlation
length of the system diverges at the threshold \cite{hunt09}.
Nonetheless, the accuracy of EMA tends to be quite good for systems
that are not in the vicinity of their percolation thresholds
\cite{koplik81}.
For the treatment of cases where $z$ is small, with significant stochastic
variation in the throats (especially when there is a high probability of
absence),
we apply a generalization of single-bond EMA in which the bond conductances
are not uniform, but instead vary depending on the extended coordinates
of the pores they connect.
The result of this approach is the 
transforming the probabilistic conductances $g_{ij}$ into deterministic
effective values $\tilde{g}_{ij}$.

Unlike Kirkpatrick's approach for the regular lattice, we do not consider the
effective conductance to be equal for every throat;
instead, we account for the dependence of a throat's properties on the 
properties and locations of the pores it connects.
Consequently, we solve for an an unknown function 
$\tilde{g}_{ij} = \tilde{g} (\bm{y}_i, \bm{y}_j)$
- effective throat conductance as a function of pore locations and attributes,
creating an effective medium which is heterogeneous and anisotropic in nature.
Note that the resulting effective medium is continuous in some sense - 
the pores are effectively smeared out over their distributions,
and the throat conductances are described by a function $\tilde{g}$.

Although for regular lattices the effective medium can have low average
coordination numbers,
the randomness in pore properties and locations in the proposed model
means that in our case the effective medium must usually 
have both large $N$ and a large coordination number $\tilde{z}$
in order to properly eliminate the local fluctuations in the original network.
In terms of the original network, the value $\tilde{z}$ is the number of pores
in what we call the \textit{locale} of pore $i$ - 
that is, the region of the medium to which pore $i$ can
(with nonzero probability) have direct connections.
Fortunately, the structure of real world porous media is generally in line
with this assumption - 
direct connections tend to occur only between nearby pores,
but there are generally many other pores in the vicinity of any given pore.

In order for the approximation proposed in this paper to be valid,
the overall statistical properties of the medium
(i.e. the distributions $f(\bm{y})$ and 
	$f(g \mid \bm{y}_1, \bm{y}_2)$)
must not change significantly within a single locale. 
This type of assumption is typically almost inevitable in the derivation
of a continuum model,
although here it is only required on relatively small scales and so
we can still account for macroscopic heterogeneity
such as spatial variation in composition, structure or porosity.
Aspects of the pores such as pore size can also be included in these
statistical properties, but it is important to note that in such a case
the locale consists of pores of varying sizes, and sharp changes in throat
statistics between connected pores are allowed only with negligible
frequency.
In real-world porous media with fractal properties (coal may serve as a notable
example) this type of approach should be entirely reasonable
\cite{vladimirov10,klimenko12};
in such cases, we expect transport to occur via a cascade process in which 
flow between large pores and much smaller ones occurs primarily via pores of
intermediate size.

The self-consistent condition of effective medium theory states that the
fluctuations over a section of the original stochastic network,
compared to its replacement in the effective medium,
must average to zero.
In our case, we consider this condition when applied to a single throat
within the network connecting pores with extended coordinates taking
the fixed values $\bm{y}_i$ and $\bm{y}_j$;
analogously to Kirkpatrick \cite{kirkpatrick73}, we express this condition as
\begin{equation}
\label{eq:ema}
	\int_0^\infty f (g \mid \bm{y}_i, \bm{y}_j)	\frac
		{g - \tilde{g} (\bm{y}_i, \bm{y}_j)}
		{g + \tilde{h} (\bm{y}_i, \bm{y}_j)}
	\, dg = 0 ,
\end{equation}
where $\tilde{h}(\bm{y}_i, \bm{y}_j)$ is the expected two-point conductance
between pores at positions $\bm{y}_i$ and $\bm{y}_j$,
\nomenclature[bh]{$h$}{Two-point conductance}
in the absence of a throat directly connecting them - this is the net flow
through the network resulting from externally forcing a unit difference in
concentration between those pores.
This two-point conductance is usually defined in terms of Green's functions
\cite{sahimi83}; a detailed discussion with exact values for many infinite
regular lattices can be seen in \cite{cserti00}.
This equation allows the throat conductance to vary with the locations of the
pores and the orientation and length of the throat, resulting in a
heterogeneous and anisotropic effective medium.
With this in mind, we term this approach the 
\textit{Nonuniform Effective Medium Approximation} (NEMA).
In its most general form, Eq.~(\ref{eq:ema}) is very difficult to solve, 
since the values for $\tilde{h}$ between any two pores depends on every other
throat in the effective medium, 
and consequently on the values of the function $\tilde{g}$,
while $\tilde{g}$ in turn depends upon $\tilde{h}$.
However, under certain conditions, we can derive approximations for
$\tilde{g}$ and $\tilde{h}$ which make the problem much more tractable. 

As long as almost all bonds in the original network have a very low
probability of being present, $\Psi \ll 1$, which naturally results
in $z \ll \tilde{z}$ and consequently (as needed)
a well-connected effective medium,
we can approximate $\tilde{g}$ by
\begin{equation}
\label{eq:g_approx}
	\tilde{g} (\bm{y}_i, \bm{y}_j) =
	\int_0^\infty	\frac
		{f(g \mid \bm{y}_i, \bm{y}_j)}
		{1/g + 1/\tilde{h}(\bm{y}_i, \bm{y}_j)}
	\,dg .
\end{equation}
For natural porous media, the assumption of $\Psi \ll 1$
should prove to be a reasonable one,
as there tend to be many pores in the vicinity of any given pore,
but relatively few of these tend to be directly connected.
For the details of the derivation of Eq.~(\ref{eq:g_approx}),
see Appendix~\ref{app:ema_g}.

It is clear that as long as an expression for the unknown function
$\tilde{h}(\bm{y}_i, \bm{y}_j)$ can be determined,
Eq.~(\ref{eq:g_approx}) specifies the values of the conductances in
the effective medium.
Fortunately, the value of $\tilde{h}$ is only important within
a pore's locale, 
since long throats should be very unlikely and have very low conductances,
resulting in negligible values of $\tilde{g}$ from Eq.~(\ref{eq:g_approx})
regardless of $\tilde{h}$.
Since the effective medium must have a large coordination number,
we can see that in general $\tilde{h} \gg \tilde{g}$
and so the presence or absence of a direct connection is
insignificant for calculating $\tilde{h}$.

To approximate the conductance between pore $i$ and pore $j$,
we consider it to consist of three conductances in series - 
two input conductances from the pores into
their locales, 
which we represent by the unknown function
$\tilde{h}_\text{in} (\bm{y})$, 
\nomenclature[in]{in}{Input (one-point) conductance}
and a third term which comes from the interconnection between
these locales.

For neighboring pores, we expect that in a porous medium the locales
of these pores should overlap significantly
\footnote{As a visual aid,
consider the case of two spheres of equal radius in three dimensions,
with the center of each sphere being located within the other;
in this case, the overlap between the spheres is at least $5/16$ of the
volume of either sphere.},
resulting in $O(\tilde{z}^2)$ connections to one another,
causing the third term to be negligible in comparison with the input
conductances.
This gives
\begin{equation}
\label{eq:h_approx}
	{\tilde{h} (\bm{y}_i, \bm{y}_j)} \approx
	\frac{1}{
	1/\tilde{h}_\text{in}(\bm{y}_i) + 
	1/\tilde{h}_\text{in}(\bm{y}_j)} ,
\end{equation}
where we approximate $\tilde{h}_\text{in}$ through connections of each pore to
the distribution of other pores in the effective medium, i.e. 
\begin{equation}
\label{eq:h_in}
	\tilde{h}_\text{in}(\bm{y}) = (N-1)
		\int_S
			f(\bm{y}^\circ) \tilde{g} (\bm{y}, \bm{y}^\circ) 
		\,d\bm{y}^\circ ,
\end{equation}
whereas integration occurs over values of $\bm{y}^\circ$ in the extended
space, with corresponding volume element $d\bm{y}^\circ$.

By substituting Eq.~(\ref{eq:h_approx}) into Eq.~(\ref{eq:g_approx})
and the result into Eq.~(\ref{eq:h_in}), and dividing both sides
by $\tilde{h}_\text{in}(\bm{y})$,  we get
\begin{equation}
	1 =
		\int_S
			\int_0^\infty	\frac
				{(N - 1) f(g, \bm{y}^\circ \mid \bm{y})}
				{1 + \tilde{h}_\text{in}(\bm{y})
					[1/g + 1/\tilde{h}_\text{in}(\bm{y}^\circ)]}
			\,dg	
		\,d\bm{y}^\circ .
\end{equation}
This equation implicitly defines the function $\tilde{h}_\text{in} (\bm{y})$,
which can then be used in Eqs.~(\ref{eq:h_approx}) and Eq.~(\ref{eq:g_approx})
to determine the values of $\tilde{g}$ within the effective medium.

Using the previous assumption that the distribution for conductances and
properties does not change significantly within a given locale, 
and since $\tilde{h}_\text{in}$ primarily depends on the values of $\tilde{g}$,
it follows that for $\bm{y}^\circ$ within the locale of $\bm{y}$,
$\tilde{h}_\text{in} (\bm{y}^\circ) \approx \tilde{h}_\text{in} (\bm{y})$.
With this approximation, the denominator of the integral becomes independent
of $\bm{y}^\circ$; in conjunction with Fubini's theorem this gives
\begin{equation}
	\frac{1}{N - 1} =
		\int_0^\infty
			\frac{1}{2 + \tilde{h}_\text{in}(\bm{y})/g}
			\int_S
				f(g, \bm{y}^\circ \mid \bm{y})
			\,d\bm{y}^\circ
		\,dg	.
\end{equation}
The inner integral evaluates to the distribution of conductances coming out of
a single pore with extended coordinates $\bm{y}$, i.e.
\begin{equation}
	f(g \mid \bm{y}) = \int_{S} f(g, \bm{y}^\circ \mid \bm{y}) \,d\bm{y}^\circ ,
\end{equation}
resulting in
\begin{equation}
	\label{eq:h_cond}
	\frac{1}{N-1} = 
		\int_0^\infty	\frac
			{f(g \mid \bm{y})}
			{2 + \tilde{h}_\text{in} (\bm{y}) / g}
		\, dg .
\end{equation}
This gives the value of $\tilde{h}_\text{in}$ for any $\bm{y}$,
which, together with Eqs.~(\ref{eq:h_approx}) and Eq.~(\ref{eq:g_approx}),
specifies the throat conductances within the effective medium.

\section{Continuum approximation} \label{sec:continuum}
\subsection{Reduced master equation} \label{sec:rme}
Applying the EMA described in the previous section
accounts for the stochastic variation in throat conductances in the original
network.
However, this still leaves a system with an equation for each individual
pore, i.e.
\begin{equation}
	\frac{d m_i}{dt} = \sum_j  \tilde{g}_{ij} (c_j - c_i) ,
\end{equation}
which is merely the master equation $(\ref{eq:master})$ but applied to
the effective medium rather than to the original random network.
In order to simplify this equation, we wish to transform it into a 
continuous form, wherein the amount of substance $m$ and concentration
$c$ are considered over the extended space rather
than within individual pores.
This is achieved by considering the ensemble of realizations of the 
pore positions and properties in the effective medium,
and then converting to continuous form by taking the limit of a summation
over discrete regions.

Assume that the extended space, $S$, is partitioned into countably many
regions $S_I$ (for $I \in \N$) such that each $S_I$ is a neighborhood of
some point in extended space, $\bm{y}_I$.
\nomenclature[iI]{$I, J$}{Region or pore group number}
Here we mean ``partition'' in the usual mathematical sense, i.e.
that the regions $S_I$ must be pairwise disjoint and cover $S$.
If we let $A_I$ be the set of pores contained within $S_I$,
\nomenclature[AA]{$A$}{Set of pore numbers} 
i.e. $A_I = \{i \mid \bm{y}_i \in S_I \}$, it follows that,
in turn, the sets $A_I$ form a partition of the set of all pores,
$A = \{ 1, \ldots, N \} $.

The total amount of substance within the region $S_I$ is given by 
\begin{equation}
	M_I = \sum_{i \in A_I} m_i ,
\end{equation}
and so, applying Eq.~(\ref{eq:master}), the transport equation for
$M_I$ in the effective medium is given by
\begin{equation} \label{eq:tr_sum}
	\frac{dM_I}{dt} = \sum_{i \in A_I} 
		\sum_{J \in \N} \sum_{j \in A_J} 
		\tilde{g}_{ij} (c_j - c_i) .
\end{equation}
Since $\tilde{g}_{ij} = \tilde{g}(\bm{y}_i, \bm{y}_j)$, and the $S_I$
are neighborhoods in extended space, it follows that as long as they
can be made sufficiently small, with $\tilde{g}$ continuous over each region
(i.e. the set of discontinuities of $\tilde{g}$ must have measure zero),
then for any  $i \in A_I, j \in A_J$ we have
\begin{equation}
	\tilde{g}_{ij} = \tilde{g}(\bm{y}_I, \bm{y}_J) .
\end{equation}
Since the value of the sum in Eq.~(\ref{eq:tr_sum}) is finite, we can
change the order of summation. If we denote the number of pores in the
region $S_I$ by $N_I$, then Eq.~(\ref{eq:tr_sum}) becomes
\begin{equation}
	\frac{dM_I}{dt} = \sum_{J \in \N} \tilde{g}(\bm{y}_I, \bm{y}_J)
		\Biggl(	N_I \sum_{j \in A_J} c_j
			- N_J \sum_{i \in A_i} c_i
		\Biggr) .
\end{equation}
As long as either (1) all pores have equivalent volume, $V_0$,
\nomenclature[i0]{$0$}{Characteristic value}
or (2) the volume
is included as a coordinate of the extended space (in which case we can set
$\xi_i^1 = \ln V_i$ without loss of generality),
the volume of each pore in a region can be considered equal, i.e. $V_i = V_I$.
We now introduce averaged concentrations within each region,
\begin{equation}
	C_I = \frac{M_I}{N_I V_I} = \frac{1}{N_I} \sum_{i \in A_I} c_i .
\end{equation}
It is important to note that these concentrations are based on the total pore
volume in each region, not the volume of physical space in each region.
With these concentrations, we can now write the master equation in terms
of the regions,
\begin{equation}
	\frac{dM_I}{dt} = \sum_{J \in \N} \tilde{g}(\bm{y}_I, \bm{y}_J)
		N_I N_J (C_J - C_I) .
\end{equation}
If we now consider the pores as being probabilistically spread over the space,
the quantities $N_I$ become the expected number of pores in each region. 
These can be approximated using the marginal distribution $f(\bm{y})$
as
\begin{equation}
	N_I = N \int_{S_I} f (\bm{y}) d\bm{y} .
\end{equation}
If we also let the regions become infinitesimal, we get the \textit{reduced
master equation},
\begin{align}
	\frac{\partial  M}{\partial t} (\bm{y}, t)
		= N^2 \int_S &\tilde{g}(\bm{y}, \bm{y}^\circ) 
				f(\bm{y}) f(\bm{y}^\circ) \notag \\
			&\times \left[ C(\bm{y}^\circ, t) - C(\bm{y}, t) \right]
			\,d\bm{y}^\circ \label{eq:rme} ,
\end{align}
where $C$ is the local concentration of the substance per unit pore volume,
\nomenclature[AC]{$C$}{Local amount of substance per unit pore volume}
and $M$ is the local concentration of the substance per unit of extended
	spatial volume.
\nomenclature[AM]{$M$}{Local amount of substance per unit extended spatial volume}

\subsection{Diffusional approximation} \label{sec:da}
In principle, the reduced master equation can be solved
directly, but a simplified form is still desirable.
Fortunately, we know from Section~\ref{sec:ema} that for any given $\bm{y}$,
$\tilde{g}$ vanishes outside its locale, while the functions $\tilde{g}$,
and $f$ must not change significantly within this locale.
Based on this, we expect $C$ to behave similarly, and so
we use power series to derive an approximation for Eq.~(\ref{eq:rme})
in the form of a diffusion-like Partial Differential Equation (PDE).

We introduce 
$\Delta \bm{y} = \bm{y}^\circ - \bm{y}
	= (\Delta y^1, \Delta y^2, \ldots)$,
\nomenclature[gdelta]{$\Delta$}{Difference between two values}
and let $\tilde{g}_2$ satisfy
\begin{equation}
		\tilde{g}(\bm{y}, \bm{y}^\circ)
	= \tilde{g}_2(\bm{y} - \bm{y}^\circ, \bm{y} + \bm{y}^\circ)
	= \tilde{g}_2(\Delta \bm{y}, 2\bm{y} + \Delta \bm{y}) .
\end{equation}
Notably, since $\tilde{g}$ is symmetric, it follows that $\tilde{g}_2$ must
\nomenclature[bg2]{$\tilde{g}_2$}{$\tilde{g}$ as a function of
	$\Delta \bm{y}$ and $2 \bm{y}$}
be an even function with respect to its first argument, $\Delta \bm{y}$.
The introduction of $\tilde{g}_2$ allows us to rewrite the reduced master
equation as
\begin{align}
	\label{eq:rme2}
	\frac{\partial M}{\partial t}
		= N^2 \int_S &\tilde{g}_2(\Delta \bm{y}, \bm{y} + \Delta \bm{y})
			f (\bm{y}) f(\bm{y} + \Delta \bm{y}) \notag \\
			&\times \left[ C(\bm{y} + \Delta \bm{y}, t) - C(\bm{y} , t) \right]
			\, d (\Delta \bm{y}) .
\end{align}
Taking power series expansions for
$\tilde{g}_2(\Delta \bm{y}, \bm{y} + \Delta \bm{y})$,
$f(\bm{y} + \Delta \bm{y})$,
and 
$C(\bm{y} + \Delta \bm{y}, t) - C(\bm{y} , t)$
yields a diffusional approximation - for detail on this
derivation, see Appendix~\ref{app:diff}.
Here we introduce
\begin{equation}
	\phi(\bm{y}) = N f(\bm{y}) V(\bm{y}) ,
\end{equation}
which means that $M(\bm{y}) = \phi(\bm{y}) C(\bm{y})$.
In the case where there are three spatial dimensions
and no additional properties, i.e. $\bm{y} = (x^1, x^2, x^3)$,
$\phi$ is in fact the local (averaged) porosity of the medium.
If $\phi$ does not depend on time, 
then 
(using the Einstein summation convention over Greek indices)
\nomenclature[pa]{$\alpha, \beta$}{Vector components
	(Einstein summation convention applies)}
the final equation takes the form
\begin{equation}
	\label{eq:de_cc}
	\phi(\bm{y}) \frac{\partial C}{\partial t}
		=  \frac{\partial}{\partial y^{\alpha}} \left[
				\phi(\bm{y}) D^{\alpha \beta} (\bm{y})
				\frac{\partial C}{\partial y^{\beta}}
			\right] ,
\end{equation}
where the elements of the diffusion tensor $D$ are given by
\nomenclature[AD]{$D$}{Diffusion coefficient}
\begin{equation}
	\label{eq:b}
	D^{\alpha \beta} (\bm{y})
		= \frac{N f(\bm{y})}{2 V (\bm{y})}
		\int_S  
			\Delta y^{\alpha}\Delta y^{\beta} %\notag \\
			\tilde{g}_2(\Delta \bm{y}, 2\bm{y})
		\, d (\Delta \bm{y}) .
\end{equation}

It is important to note that while $C$ is not conserved,
Eq.~(\ref{eq:de_cc}) is conservative with respect to $M = \phi C$,
as should be the case for transport with no sources or sinks.
The above integral is assumed to converge; any case where it does not
would likely be a case of anomalous diffusion, which cannot be represented
by the above equations.
It can be seen that this equation takes a form very similar to the 
Fokker-Planck (Kolmogorov forward) equation, which governs the PDF of an 
It\={o}-type random walk.
However, the Kolmogorov forward equation differs in that the diffusion
coefficient is differentiated twice with respect to $\bm{y}$.

\section{Results} \label{sec:results}
In this section, the accuracy of the approximation methodology described in
Sections~\ref{sec:ema} and \ref{sec:continuum} is evaluated for networks
with different types of distributions for conductances by comparison
against numerical simulation results.
For convenience, we abbreviate the overall methodology as NEMA-DCA
(Nonuniform Effective Medium Approximation and
Diffusional Continuum Approximation).
As a further point of comparison, a more direct application of the diffusional
approximation is given, in which each conductance is first replaced by its mean
or expected value - this is precisely the simplified approach suggested at the
very beginning of Section~\ref{sec:ema}; we term this approach the
Averaged Diffusional Approximation (ADA).
Note that the use of the effective medium approximation by itself (i.e. without
a continuum approximation) does not directly yield useful results,
since the resulting transport equations are not less complex than the original
master equation.

To simplify the mathematics, we consider cases where the conductances do not
vary with space or direction, but have different probability distributions
with respect to the throat length.
Three characteristic types are considered for the distribution of nonzero
conductances within the medium:
\begin{itemize}
	\item A single, discrete value (Section~\ref{sec:svd}).
	\item  A continuous distribution of conductances
		(Section~\ref{sec:diid}).	
		For comparison, this distribution is chosen
		to have the same shape as in the previous case,
		but independently of throat length.
	\item Conductances inversely proportional to throat length
			(Section~\ref{sec:ild}).
\end{itemize}

\subsection{Simulation parameters}
To simplify the calculations, the pore networks are considered to be contained
within an $n$-dimensional cube having side length $w$,
\nomenclature[bw]{$w$}{Side length for cubic sample}
in which the pores are independently distributed with a uniform distribution.
The number of pores is set to $N = 10000$, and the pores are assumed to be
identical in terms of their additional properties $\bm{\xi}$, resulting in
all pores having the same volume, $V_0$.
As such, the extended space can be considered to consist of 
only the spatial dimensions, $\bm{y} = \bm{x}$.
Consequently, the pore network is contained within a hypercube of
$n$-volume $w^n$, and within this cube $f(\bm{x}) \equiv w^{-n}$,
resulting in constant porosity
\begin{equation}
	\phi(\bm{y}) = \phi_0 = N V_0 / w^n ,
\end{equation}

For further simplification, we consider a case where the probability
distribution for the conductance of a throat is determined only by the
quantity $r_{ij} = \norm{\bm{x}_i - \bm{x}_j}$,
\nomenclature[br]{$r$}{Distance between pores}
which is the distance between the pores - this is directly related
to the throat length, though in general they are not equal since
the pores are not single points and the throats may be curved.
Consequently,
$f (g \mid \bm{x}_i, \bm{x}_j) = f (g \mid r_{ij})$,
and $\tilde{g}_2 (\Delta \bm{x}, 2\bm{x}) = \tilde{g}(r)$.
As a result of this, the simulated networks can be described
by a single diffusion coefficient which is independent of position and
orientation - that is, they are homogeneous and isotropic with respect
to macroscopic diffusivity.

In order to localize the connections within the medium,
we define the throat probability function as a rectangular function of the
inter-pore distance,
\nomenclature[im]{$m$}{Maximum value}
\begin{equation}
	\Psi (r) = 
	\begin{cases}
		0, &\text{if } r > r_m \\
		p, &\text{if } r \leq r_m ,
	\end{cases}
\end{equation}

From Eq.~(\ref{eq:z}), we can see that the expected coordination number for
any given pore is independent of its location, and takes the form
\begin{equation} 
	z
		= \frac{pN K_n}{w^n} \int_0^{r_m} r^{n-1} \, dr
		= \frac{pN K_n r_m^n}{nw^n} ,
\end{equation}
where $K_n$ is the ``surface area'' of an $n$-sphere of unit radius.
\nomenclature[AKn]{$K_n$}{Surface area of an $n$-sphere of unit radius}
For convenience, we select as a characteristic distance $r_0$
the root mean squared distance between connected pores in the effective
medium, which in this case is described by the equation
\begin{equation}
	\frac{r^2_0}{n} = \frac{r^2_m}{n+2} .
\end{equation}

\subsection{Analytical methods}
Given the simulation parameters above, application of the EMA described in
Section~\ref{sec:ema} results in a new network whose average coordination
number is $\tilde{z} = z / p$, with no throats connecting pores further than
$r_m$ apart, i.e. $\tilde{g} (r) = 0$ for $r > r_m$.
On the other hand, for $r \leq r_m$, there is a single value of 
$\tilde{h}_\text{in}$ that is constant throughout the medium,
and Eq.~(\ref{eq:h_cond}) takes the form
\begin{equation}
	\label{eq:h_cond_r}
	\frac{1}{N-1} = 
		\int_0^\infty	\frac
			{f (g)}
			{2 + \tilde{h}_\text{in} / g}
		\, dg .
\end{equation}
For $r \leq r_m$, we have $\tilde{h} = \tilde{h}_\text{in} / 2$ and so
Eq.~(\ref{eq:g_approx}) takes the form
\begin{equation}
	\label{eq:g_approx_r}
	\tilde{g} (r) = \int_0^\infty	\frac
		{f(g \mid r)}
		{1/g + 2/\tilde{h}_\text{in}}
	\,dg ,
\end{equation}
while the diffusional approximation of Eq.~(\ref{eq:de_cc})
takes the simplified homogeneous and isotropic form
\begin{equation}
	\frac{\partial C}{\partial t} (\bm{x}, t)
		= D \nabla^2 C(\bm{x}, t) ,
	\label{eq:da_std}
\end{equation}
in which the diffusion coefficient is
\begin{equation}
	D
	= \frac{\tilde{z}}{2 V_0 r_m^n} \int_0^{r_m}
		r^{n+1} \tilde{g}(r)
		\, dr .
	\label{eq:d_th}
\end{equation}

\subsection{Numerical method}
In order to calculate numerical results for comparison with the
derived equations, a relatively simple approach has been used.
The graph is constructed as a symmetric $N \times N$ sparse matrix, in typical
fashion for a weighted graph - throat conductances are represented by the
nonzero values.
In a modified form, this matrix can represent the master equation for every
pore in the system as a matrix equation, which can be solved subject to
appropriate boundary conditions.

In these examples, steady-state solutions to the master equation are
considered, in which the boundary conditions are specified by 
setting pores on opposing sides - specifically,
within distance $\epsilon$ of opposing faces of the cube - are fixed
at different levels of concentration,
and the resulting flow rate $Q$ is calculated using 
the aforementioned matrix method.
\nomenclature[AQ]{$Q$}{Net flow rate through the medium}
This allows simulated diffusion coefficients to be calculated as
\begin{equation}
	D_\text{sim} = \frac{Q w (w - 2 \epsilon)}{N V_0 \Delta C} ,
\end{equation}
where $\Delta C$ is the difference in concentration between the
two sides.

The following sections will compare the numerically derived
diffusion coefficients to analytical results using ADA (labeled
as $D_a$)
\nomenclature[ia]{$a$}{Using ADA}
and using NEMA-DCA ($D_\text{em}$).
\nomenclature[iem]{em}{Using NEMA-DCA}
In order to display the nontrivial dependence of the diffusion coefficients
on $z$, the diffusion coefficients shown in the figures are normalized to the
dimensionless form
\nomenclature[zh]{Hat (e.g. $\hat{D}$)}{Normalized dimensionless value}
\begin{equation}
	\hat{D} = D \frac{n V_0 }{g_0 r_0^2}
\label{eq:norm}
\end{equation}
where $g_0$ is a characteristic conductance value.
Although $g_0$ is defined differently in some of the cases below,
in each case the scaling factor ${n V_0}/{g_0 r_0^2}$ is the same
for both the analytical and numerical methods.

\subsection{Single-valued distribution} \label{sec:svd}
In this case, a class of networks is evaluated in which the conductance values,
when present, always have a fixed conductance value, $g_0$.
As stated previously, each throat has probability $p = z /\tilde{z}$ of being
present if the distance between the pores is no greater than $r_m$, and zero
probability of being present otherwise.

Applying the ADA method would use $\bar{g}(r) = (z / \tilde{z}) g_0$;
implementing this instead of $\tilde{g}$ in Eq.~(\ref{eq:d_th}) predicts 
a diffusion coefficient of
\begin{equation}
	D_a
	= \frac{z}{2}  \frac{g_0 r_0^2}{n V_0}.
\end{equation}

On the other hand, for the NEMA-DCA method, 
Eq.~(\ref{eq:h_cond_r}) gives
\begin{equation}
	\tilde{h}_\text{in} = g_0 (z - 2) .
\end{equation}
Applying Eq.~(\ref{eq:g_approx_r}) for $r \leq r_m$ leads to 
\begin{equation}
	\tilde{g}(r) =
	\frac{z-2}{\tilde{z}}
	g_0 ,
\end{equation}
which, when used in Eq.~(\ref{eq:d_th}), gives
\begin{equation}
	D_\text{em}
	= \frac{z-2}{2} \frac{g_0 r_0^2}{n V_0} .
\end{equation}

\begin{figure}[t!]
\includegraphics{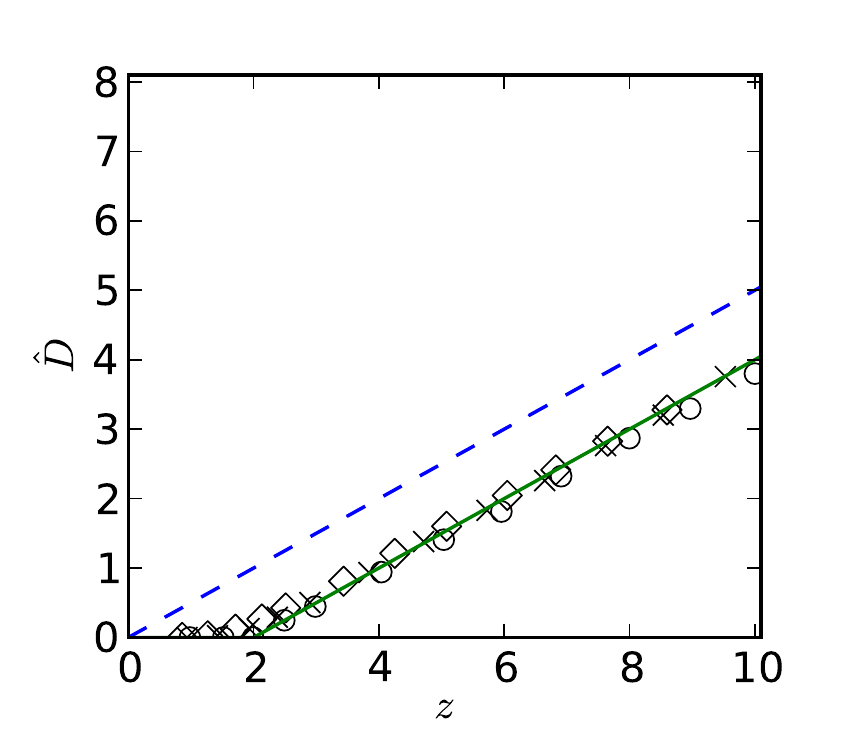}
\caption{(Color online) Results for the single-valued conductance distribution.
The circle, cross, and diamond markers are the results of simulations for
$n=1$, $n=2$, and $n=3$ respectively, the dashed line is derived from ADA,
and the solid line is derived from NEMA-DCA.}
\label{fig:sv}
\end{figure}

A comparison of simulation results in 1, 2 and 3 dimensions
to the predicted diffusion coefficients
for the single-valued conductance distribution is given in Fig.~\ref{fig:sv}. 
Note that the axes, $z$ and $\hat{D}$, are both dimensionless.
The graph demonstrates the accuracy of NEMA-DCA for varying numbers of physical
dimensions, $n$, and varying values of $z$. While ADA predicts the
asymptotic behavior as $z \to \infty$, its relative error is rather high at
lower values of $z$. On the other hand, NEMA-DCA performs quite well for
$z \gtrsim 2$. Nonetheless, neither approach correctly predicts the behavior
near the percolation limit, nor the percolation limit itself - this is to
be expected, as EMA methodologies in general fail in approximating
percolation.

Although our networks take an irregular form, the results shown in
Fig.~\ref{fig:sv} are quite similar to those for 
Kirkpatrick's simple regular network \cite{kirkpatrick73}.
The predictions of NEMA-DCA reflect the tendency for low coordination numbers to
significantly reduce transport coefficients in physical porous media,
as compared to the predictions of a simple diffusional approximation.

\subsection{Distance-independent inverse distribution} \label{sec:diid}
As before, we consider $n=1$, with no throats for pores further than $r_m$
apart, while throats covering shorter distances have a fixed probability
$z / \tilde{z}$ of being present.
However, in this case, when a throat is present its conductance is 
determined (independently of $r$) by the distribution
\nomenclature[bk]{$k$}{Constant in inverse conductance distribution}
\begin{equation} \label{eq:inv_dist}
	f_G^*(g) = \frac{k} {r_m} g^2
\end{equation}
where $k$ is a constant.
As mentioned earlier, the star superscript indicates that this
distribution is conditional on a throat actually being present,
i.e. $G>0$.
For this example, we define the characteristic conductance as $g_0 = k / r_0$.
Notably, ADA would fail entirely in this case, as $\bar{g} = \infty$ and so
\begin{equation}
	D_a = \infty .
\end{equation}

\begin{figure}[!t]
\includegraphics{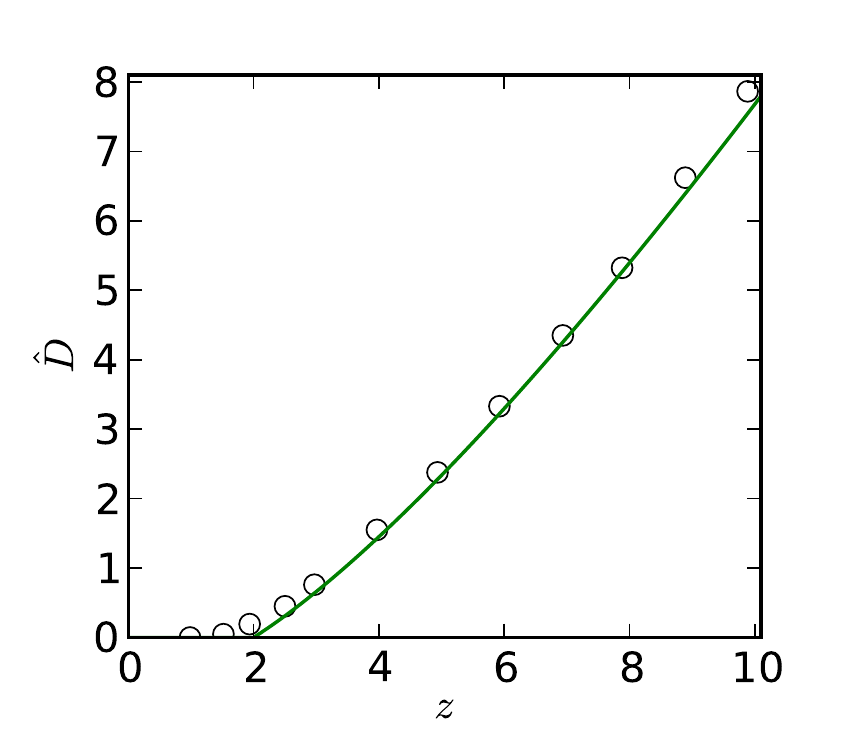}
\caption{(Color online) Results for the distance-independent inverse
conductance distribution, all for $n=1$.
The circle markers are simulation results, while the solid line is derived
from NEMA-DCA.
No line is shown for ADA, as it predicts infinite coefficients in this case.}
\label{fig:ii}
\end{figure}

However, the application of NEMA removes this singularity.
By using Eq.~(\ref{eq:inv_dist}) in Eq.~(\ref{eq:h_cond_r}), we get
\begin{equation}
	\label{eq:h_in_inv}
	\frac{r_m}{k z} = 
		\int_{k / r_m}^\infty	\frac
			{1}
			{2 g^2 +  \tilde{h}_\text{in} g}
		\, dg .
\end{equation}
If we define the function $B$ implicitly by 
\nomenclature[AB]{$B$}{Function defined by Eq.~(\ref{eq:B})}
\nomenclature[bs]{$s$}{Dummy variable in the definition of $B$}
\begin{equation}
1 + B(s) = \exp \left( \frac{2B(s)}{s} \right)
\label{eq:B} ,
\end{equation}
then $\tilde{h}_\text{in}$ can be expressed as
$\tilde{h}_\text{in} = 2 k B(z) / r_m$.

However, in this case Eq.~(\ref{eq:g_approx_r}) becomes
\begin{equation}
	\label{eq:g_2}
	\tilde{g}(r) = \frac{2 k B(z)}{r_m \tilde{z}} ,
\end{equation}
for $r \leq r_m$.
Note that, as in the single-valued case, this value does not depend on $r$,
and so every throat in the effective medium has the same conductance.

Using Eq.~(\ref{eq:g_2}) in Eq.~(\ref{eq:d_th}) we derive the diffusion
coefficient 
\begin{equation}
	D_\text{em}
	= \frac{B(z)}{\sqrt{3}} \frac{g_0 r_0^2}{V_0} .
\end{equation}

A comparison of simulation results to the predicted diffusion coefficients
for the distance-independent inverse conductance distribution is
given in Fig.~\ref{fig:ii}.
Once again, NEMA-DCA performs quite well, though it does not predict the
percolation threshold accurately.
On the other hand, ADA in this case is only valid for infinite $z$, as
it predicts infinite diffusion coefficients.

Although this example presents an extreme case, it serves to illustrate
the use of effective medium techniques in order to provide a realistic
account for the effects of sparse high-throughput connections within
irregular porous media.
Such connections can be present in physical porous media, but it is clear
that the overall throughput would nonetheless be bottlenecked by the 
smaller connections, as is predicted by NEMA-DCA for lower values of $z$.

\begin{figure}[t!]
\includegraphics{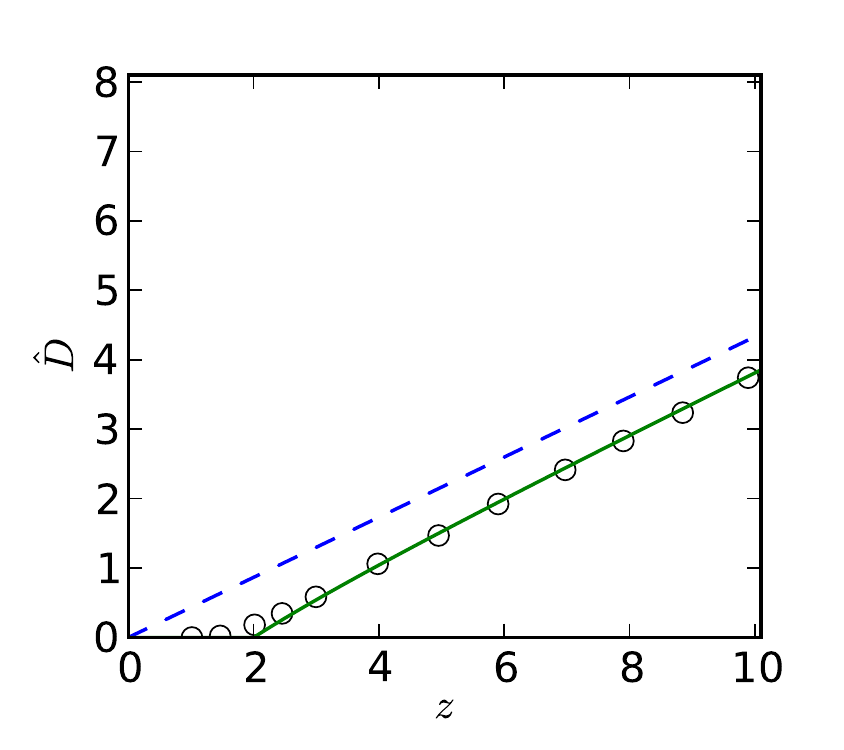}
\caption{(Color online) Results for the inverse-distance conductance
distribution, all for $n=1$.
The circle markers are simulation results, the dashed line is derived from ADA,
and the solid line is derived from NEMA-DCA.}
\label{fig:ir}
\end{figure}

\subsection{Inverse-distance distribution} \label{sec:ild}
For this example, we once again have throats being present with probability
$z / \tilde{z}$ if $r \leq r_m$, and as in the previous example we
select $n = 1$.
However, in this case, whenever such a throat is indeed present, its
conductance takes the value $g = k / r$.
For simplicity and ease of comparison to the previous case, we define the
characteristic conductance as $g_0 = k / r_0$.

In this case, the ADA approach would use
$\bar{g}(r) = (z / \tilde{z}) (k/r)$, which when used in 
Eq.~(\ref{eq:d_th}) gives
\begin{equation}
	D_a
	= \frac{\sqrt{3}z}{4} \frac{g_0 r_0^2}{V_0}
\end{equation}
To apply NEMA-DCA, we require a PDF for the conductances.
Since $f^*(r) = 1 / r_m$, it follows that
\begin{equation}
	f^*(g) = \frac{k} {r_m} g^2.
\end{equation}
Notably, this is, in fact, the same distribution of conductances as in the
previous example, and the inter-pore distances $r$ also retain the same distribution 
as before,
but this case is quite different from the previous one due to the correlation between
the conductance and inter-pore distance.
Consequently, if the dependence of conductance on distance were ignored,
NEMA-DCA would make the same prediction here as in the previous case.

Since the conductance distribution is the same, Eq.~\ref{eq:h_cond_r} takes
an identical form to that in the previous case, and so once again we have
$\tilde{h}_\text{in} = 2 k B(z) / r_m$, while the function $B(z)$ is 
determined by Eq.~(\ref{eq:B}).
With Eq.~(\ref{eq:g_approx_r}) for $r \leq r_m$ this gives
\begin{equation}
	\tilde{g}(r) = 
	\frac{z / \tilde{z}}
		{r/k + r_m / [k B(z)]} ,
\end{equation}
and, from Eq.~(\ref{eq:d_th}),
\begin{align}
	D_\text{em}
	&= \frac{kz}{2 V_0 r_m} \int_0^{r_m}
		\frac{r^2} {r + r_m / B(z)} \, dr , \notag \\
	&= \frac{\sqrt{3}}{4}
		\left[
			z - 2 \frac{z-2}{B(z)}
		\right]
		\frac{g_0 r_0^2}{V_0} 
\end{align}

A comparison of simulation results to the predicted diffusion coefficients
for the inverse-distance conductance distribution is given in
Fig.~\ref{fig:ir}.
As can be seen, NEMA-DCA is very accurate except near the percolation limit,
while ADA predicts the asymptotic behavior and is accurate only for
larger $z$.
Like the first example, this example illustrates the need to account
for the effects of limited interconnectedness in order to achieve
a realistic model.

\subsection{Overall discussion}
In contrasting the last two cases, with differing inverse distributions,
it is clear from Figs.~\ref{fig:ii} and \ref{fig:ir}
that the results are very different - the radius-independent distribution
gives a higher diffusion coefficient by a factor  of $2$ at $z=10$,
and it is clear from the shape of the curve that this divergence
persists for increasing $z$.
This shows the importance of properly representing various property
correlations, especially between pores and throats, in the porous medium.
Such correlations are the only difference between the cases in
Fig.~\ref{fig:ii} and Fig.~\ref{fig:ir}, and it is clear that these kinds
of correlations are an essential aspect of real-world porous media.
Unlike NEMA, a more typical EMA (e.g. Kirkpatrick \cite{kirkpatrick73})
would not be able to account for this type of behavior.

However, it is interesting to note that in each case, the NEMA predicts a
percolation limit at $z=2$, just as in Kirkpatrick's paper
\cite{kirkpatrick73} - this seems to be a universal property of single-bond
effective medium approximations, regardless of whether the network in 
question is regular or irregular in nature.
Notably, as is demonstrated by the second example, the behavior away from
$z=2$ does not need to be linear.
Although the numerical results show that the behavior very close to the
true percolation limit is not correctly predicted,
as is generally the case for effective medium approximations,
it can be seen that the NEMA-DCA prediction starts to match the numerical
results very closely at a short distance from the percolation limit.

Moreover, the general interdisciplinary tool developed in this paper can
be applied to Fickian diffusion, Knudsen diffusion, and
incompressible Newtonian Stokes flow;
example throat conductance expressions for some simple throat geometries are
presented in Table 1 of Appendix \ref{app:trans}.

\section{Conclusions}
In this paper, we have presented a general analytical methodology for analyzing 
various transport phenomena in irregular porous media.
We have developed a generalized effective medium approximation which
allows the conductances of the throats in the medium to vary depending on the
properties and positions of the pores they connect.
This is used in combination with a continuum approximation in the form of
a diffusion-like equation to derive an explicit analytical expression for
the transport coefficients, allowing for an efficient model of transport
that retains the ability to account for microscopic effects.

These analytical expressions have been tested against several numerical
simulations with different types of distributions for throat properties,
demonstrating very good accuracy for the proposed
methodology, except in the vicinity of the percolation limit.
Furthermore, the results demonstrate the ability of this method to account
for several physical effects, including reduced transport due to low
coordination numbers, and the effects of correlations between local
properties of the medium.

\begin{acknowledgments}
The authors thank Dr. A. Skvortsov for useful discussions.
K.H. acknowledges the support provided by the Australian Research Council. 
\end{acknowledgments}

\appendix
\section{Expressions for throat conductances}\label{app:trans}
Recalling that the master equation
\begin{equation}
	\frac{d m_i}{dt} = \sum_{j=1}^N q_{ji} = \sum_{j=1}^N g_{ij} (c_j - c_i)
\end{equation}
applies to a variety of different types of transport, 
Table~\ref{table:ex_g} gives examples of throat conductances for a variety
of different cases.

\begin{table}[!ht]
  \begin{tabular}{l c c r}
		Case~ & \multicolumn{2}{c}{Expression for $g$} & Ref. \\
		& Small orifices & Long, narrow throats\\
		& ($L=0$) & ($L \gg a)$ \\
    \hline
		A & $2 D a $ & $D \pi a^2 / L$ & \cite{dagdug03} \\
		B & $\bar{v} \pi a^2 / 4$ & $2 \bar{v} \pi a^3 / (3L)$ &\cite{present58} \\
    C & $a^3 / (3 \mu)$ & $\pi a^4 / (8 \mu L)$ & \cite{weissberg62} \\
    D & ----- & $\sigma \pi a^2 / L$ \\
  \end{tabular}
	\caption{Expressions for throat conductance $g$ for various
		transport phenomena. 
		Where applicable, 
		$a$ is the radius of the throat,
		$L$ is the length of the throat, 
		$D$ is the diffusion coefficient,
		$\bar{v}$ is the mean molecular speed,
		$\mu$ is the fluid viscosity,
		and $\sigma$ is the electrical conductivity of the material.}
	\label{table:ex_g}
\end{table}

\nomenclature[ba]{$a$}{Throat radius}
\nomenclature[AL]{$L$}{Throat length}
\nomenclature[bvb]{$\bar{v}$}{Mean molecular speed}
\nomenclature[gmu]{$\mu$}{Fluid viscosity}
\nomenclature[gsigma]{$\sigma$}{Electrical conductivity}

The cases considered in Table~\ref{table:ex_g} are:
\begin{enumerate}[(A)]
	\item \textbf{Fickian Diffusion.}
	\item \textbf{Knudsen Diffusion.}
	\item \textbf{Incompressible Newtonian Stokes flow.}
		Since the flow is steady, there is no time derivative term; 
		$c_i$ is interpreted as the pressure in each pore, while
		$q_{ij}$ is the volumetric flow rate through the throat from
		pore $i$ to pore $j$.
	\item \textbf{Flow of electrical current.}
		Once again, the flow is steady;
		$c_i$ is the voltage at node (pore) $i$, and
		$q_{ij}$ is the current flowing through the branch (throat)
		from node $i$ to node $j$.
		Note that in the case of electrical transport,
		the ``pores'' and ``throats'' usually consist of the solid
		matrix rather than the void space.
\end{enumerate}

\section{\texorpdfstring
	{Approximation for $\tilde{g}$}
	{Approximation for g tilde}
}\label{app:ema_g}
Starting with the effective medium criterion [Eq.~(\ref{eq:ema})],
\begin{equation}
	\int_0^\infty f (g \mid \bm{y}_i, \bm{y}_j)	\frac
		{g - \tilde{g} (\bm{y}_i, \bm{y}_j)}
		{g + \tilde{h} (\bm{y}_i, \bm{y}_j)}
	\, dg = 0 ,
\end{equation}
we apply Eq.~(\ref{eq:dist_g}), resulting in
\begin{equation}
(1 - \Psi) \frac
	{\tilde{g}}
	{\tilde{h}}
= \Psi 
	\int_0^\infty f^*(g \mid \bm{y}_i, \bm{y}_j) \frac
		{g - \tilde{g}}
		{g + \tilde{h}}
	\, dg ,
\end{equation}
where $\Psi$, $\tilde{g}$ and $\tilde{h}$ are, as before, functions of 
$\bm{y}_i$ and $\bm{y}_j$ - the arguments have been dropped for compactness.
This can be rearranged to give
\begin{equation}
	\label{eq:g_exact}
	\tilde{g} = \frac{\Psi}{\chi}
		\int_0^\infty	\frac
			{f^*(g \mid \bm{y}_i, \bm{y}_j)}
			{1/g + 1/\tilde{h}}
		\, dg ,
\end{equation}
where 
\begin{equation}
	\chi = 1 - 
		\Psi \left[1 -
			\int_0^\infty \frac
				{f^*(g \mid \bm{y}_i, \bm{y}_j)}
				{1 + g/\tilde{h}}
			\,dg
		\right] .
\end{equation}
Since $f^*$ is a PDF and the conductances must take positive values,
the integral within the brackets is
bounded below by $0$ and above by $1$, and so
$1 - \Psi \leq \chi \leq 1$.
For $\Psi \ll 1$, we get $\chi \sim 1$
and hence Eq.~(\ref{eq:g_exact}) can be written as
\begin{equation}
	\tilde{g} (\bm{y}_i, \bm{y}_j) = 
	\int_0^\infty	\frac
		{f(g \mid \bm{y}_i, \bm{y}_j)}
		{1/g + 1/\tilde{h}(\bm{y}_i, \bm{y}_j)}
	\,dg ,
\end{equation}
since the $\delta(g)$ component of the integral vanishes due to the $1/g$
term in the denominator.
This completes the derivation of Eq.~(\ref{eq:g_approx}).

\begin{widetext}
\section{Derivation of the diffusional approximation} \label{app:diff}
Based on the assumption in Section~\ref{sec:ema} that the network has
no long-range connectivity and its statistical properties do not change
significantly within any locale, we expect that
power series expansions should offer reasonable approximations
for $f$ and $\tilde{g_2}$, and, based on the form of 
Eq.~(\ref{eq:rme2}), for $C$.
The power series expansions for
$C(\bm{y} + \Delta \bm{y}, t) - C(\bm{y} , t)$,
$f(\bm{y} + \Delta \bm{y})$,
and
$\tilde{g}_2(\Delta \bm{y}, \bm{y} + \Delta \bm{y})$,
are as follows:
\begin{equation}
	C(\bm{y} + \Delta \bm{y}, t) - C(\bm{y}, t)
		= \Delta y^{\alpha} \frac
			{\partial C}
			{\partial y^{\alpha}}
			(\bm{y} , t)
		+ \frac{1}{2} \Delta y^{\alpha} \Delta y^{\beta} \frac
			{\partial^2 C}
			{\partial y^{\alpha} \partial y^{\beta}}
			(\bm{y} , t) + \ldots ,
\label{eq:c_pse}
\end{equation}
\begin{equation}
f(\bm{y} + \Delta \bm{y})
	= f(\bm{y}) 
	+ \Delta y^{\alpha} \frac
			{\partial f}
			{\partial y^{\alpha}}
			(\bm{y})
	+ \ldots \, ,
\label{eq:f_pse}
\end{equation}
and
\begin{equation}
\tilde{g}_2(\Delta \bm{y}, 2\bm{y} + \Delta \bm{y}) 
	= \tilde{g}_2(\Delta \bm{y}, 2\bm{y}) +
		\frac{1}{2} \Delta y^{\alpha} \frac
			{\partial \tilde{g}_2}
			{\partial y^{\alpha}}
		 (\Delta \bm{y}, 2\bm{y})
		+ \ldots \, .
\label{eq:g_pse}
\end{equation}
Using these expansions while discarding third- and fourth-order terms
allows the reduced master equation to be expressed as
\begin{equation}
	\frac{\partial M}{\partial t}
		= \frac{N^2}{2} \int_S \Delta y^{\alpha} \Delta y^{\beta} \left(
			f^2 \tilde{g}_2 \frac{\partial^2 C}
			                     {\partial y^{\alpha}\partial y^{\beta}}
			+ f^2 \frac{\partial \tilde{g}_2}{\partial y^\alpha}
				\frac{\partial C}{\partial y^\beta}
			+ 2f \frac{\partial f}{\partial y^\alpha} \tilde{g}_2
				\frac{\partial C}{\partial y^\beta}
		\right) \, d (\Delta \bm{y}) ,
\label{eq:rme_long}
\end{equation}
\end{widetext}
where the arguments for the functions are dropped for compactness;
as before,
$f$ and its derivatives are evaluated at $\bm{y}$, $\tilde{g}_2$ and its
derivatives are evaluated at $(\Delta \bm{y}, 2\bm{y})$, and $C$ and its
derivatives are evaluated at $(\bm{y}, t)$.

Note that the first-order (drift) terms in this equation are zero.
This follows from the fact that
$f(\bm{y})$ and $C(\bm{y}, t)$ are independent of $\Delta \bm{y}$
while $g_2$ is even with respect to $\Delta \bm{y}$,
resulting in 
\begin{equation}
	 \int_S \Delta y^{\alpha} \tilde{g}_2 \, d (\Delta \bm{y}) = 0 ,
\end{equation}
Using the product rule, as well as the aforementioned independence,
Eq.~(\ref{eq:rme_long}) can be expressed as
\begin{equation}
	\frac{\partial M}{\partial t}
		= \frac{\partial}{\partial y^{\alpha}} \left[ \frac{N^2}{2} 
				f^2 \int_S \Delta y^{\alpha} \Delta y^{\beta}
				\tilde{g}_2 \, d (\Delta \bm{y})
				\frac{\partial C}{\partial y^{\beta}}
			\right] ,
\end{equation}
completing the derivation of the diffusional approximation.

%\bibliography{applications,articles,books,recons,emt}
%

\end{document}